# Potential of the three-terminal heterojunction bipolar transistor solar cell for space applications

Antonio Martí
*Instituto de Energía Solar*
*Universidad Politécnica de Madrid*
Madrid, Spain
ORCID: 0000-0002-8841-7091

Pablo García-Linares
*Instituto de Energía Solar*
*Universidad Politécnica de Madrid*
Madrid, Spain
p.garcia-linares@upm.es

Marius Zehender
*Instituto de Energía Solar*
*Universidad Politécnica de Madrid*
Madrid, Spain
marius.zehender@alumnos.upm.es

Simon A. Svatek
*Instituto de Energía Solar*
*Universidad Politécnica de Madrid*
Madrid, Spain
simon.svatek@upm.es

Irene Artacho
*Instituto de Energía Solar*
*Universidad Politécnica de Madrid*
Madrid, Spain
irene.artacho@ies-def.upm.es

Ana B. Cristóbal
*Instituto de Energía Solar*
*Universidad Politécnica de Madrid*
Madrid, Spain
anabel.cristobal@ies-def.upm.es

José R. González
*ESTEC*
*European Space Agency*
Noordwijk, The Netherlands
Jose.Ramon.Gonzalez@esa.int

Carsten Baur
*ESTEC*
*European Space Agency*
Noordwijk, The Netherlands
Carsten.Baur@esa.int

Íñigo Ramiro
*Institut de Ciències Fotòniques*
*The Barcelona Institute of Science and Technology*
Barcelona, Spain
inigo.ramiro@icfo.eu

Federica Cappelluti
*Department of Electronics and Telecommunications*
*Politecnico di Torino*
Torino, Italy
federica.cappelluti@polito.it

Elisa Antolín
*Instituto de Energía Solar*
*Universidad Politécnica de Madrid*
Madrid, Spain
elisa.antolin@upm.es

*Abstract*— Multi-terminal multi-junction solar cells (MJSC) offer higher efficiency potential than series connected (two-terminal) ones. In addition, for terrestrial applications, the efficiency of multi-terminal solar cells is less sensitive to solar spectral variations than the two-terminal series-connected one. In space, generally, cells are always illuminated with AM0 spectrum and no impact is expected from spectral variations. Still, in space, the multi-terminal approach offers some advantages in comparison with the series-connected architecture approach derived from a higher end of life (EOL) efficiency. In this work we review the potential of multi-terminal solar cells for achieving extended EOL efficiencies with emphasis in the potential of the three-terminal heterojunction bipolar transistor solar cell, a novel multi-terminal MJSC architecture with a simplified structure not requiring, for example, tunnel junctions.

## I. Introduction

Two-terminal multi-junction solar cells (MJSC) comprise series connected subcells that show an increasing sensitivity to spectral variations as the number of junctions increases. In [1] some of us showed and discussed that, implementing independent connections to a four junction (4J) MJSC offers an annual energy efficiency gain of +4.5 points while, going from 4J to five-junction (5J) series-connected cell, only offers +1.3 efficiency points. In space, however, solar cells are expected to be illuminated by a constant spectrum (AM0) so that this advantage of the multi-terminal approach does not apply. We think, however, that a multi-terminal approach still has advantages in space applications in terms of extending the end of life (EOL) efficiency of the solar cells by minimizing the impact on the cell total efficiency of the degradation of one of the subcells in the stack when these are independently connected with respect to the two-terminal case.

## II. Description of the Solar Cells Studied in this Work

We will study the efficiency in space conditions of the three types of solar cells illustrated in Fig. 1 and described next.

The first type of solar cell we will consider is the conventional three-junction solar cell connected in series leading to a two-terminal device. For short, we will designate this cell as 3J-2T. The top subcell is made of InGaP, the middle subcell of GaAs and the bottom subcell of Ge. Notice that, in this structure, subcells are connected through tunnel junctions and that, due to the series connection between the subcells, the electrical current that circulates across all the subcells is the same. Being this one of the most conventional solar cells used in space applications, its analysis is useful in order to perform as baseline for the other cases.

The second type of solar cell, that we will designate as 3J-6T, consists of the same basic semiconductor structure than in the 3J-2T case, but in which we have extracted two terminals from each subcell which, when globally considered, lead us to a six-terminal device. Notice that, in this case, in order to extract two independent terminals from each cell, the internal connection between the cells must be electrically isolating and not by means of tunnel junctions. Obviously, in this structure, the operation point of each subcell can be fixed independently from each other.

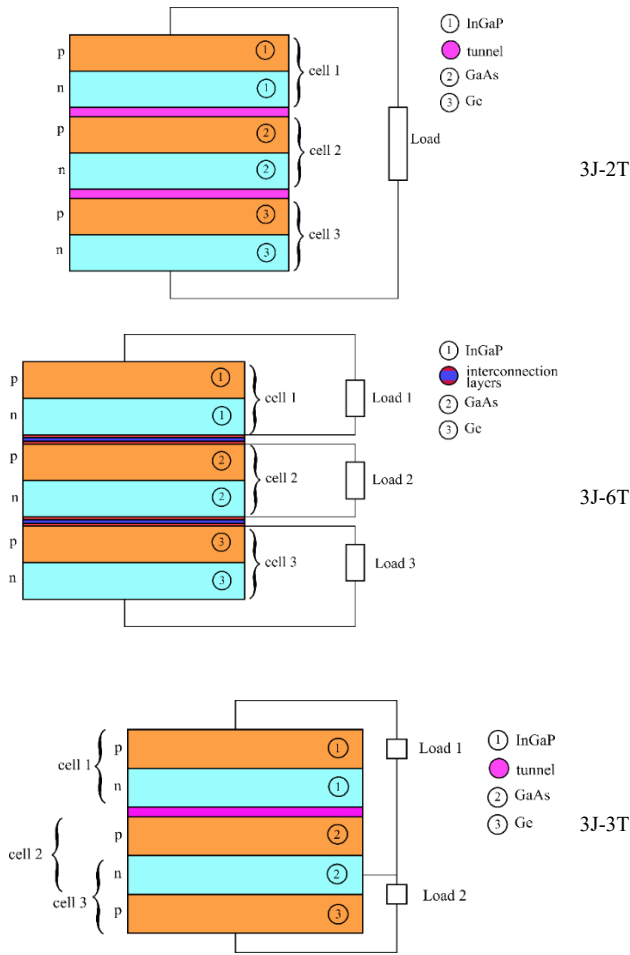

Fig. 1. Multi-terminal and multi-junction solar cell structures studied in this work: 3J-2T: three-junction solar cell connected in series leading to a two-terminal device (conventional solar cell); 3J-6T: three-junction solar cell independently connected leading to a six-terminal device; 3J-3T: three-junction solar cell following a three-terminal heterojunction bipolar transistor approach leading to a three-terminal device.

Finally, the third type of multi-junction solar cell that we shall consider, groups the middle and bottom cell of the conventional 3J-2T solar cell structure into a 3T-HBTSC (heterojunction bipolar transistor solar cell). This has the advantage of simplifying the internal semiconductor structure of the cell (by removing, for example, one of its tunnel junctions) although at the cost of adding one additional terminal. On the other hand, the 3T-HBTSC is connected in series with the same top subcell of the 3J-2T. The motivation for making this choice about where do we take out the extra terminal comes from the desire of taking advantage of the extra current generated by the Ge bottom cell which becomes limited in the conventional 3T-2J cell, because of the series connection.

III. MODEL

We explain in this section the simplified model we have used to calculate the efficiencies of the solar cells we have described in the previous section. The parameters used in the simulations are collected in Table I.

In all cases, we have assumed the cells to operate in the radiative limit with the few modifications that we shall describe later. Therefore, to calculate the current-voltage characteristic of the subcells we have used Shockley and Queisser (S&Q) [2] detailed balance model, refined and described with more detail in [3] and adapted for multi-junction solar cells in [4]. It is convenient to remember that, by using this model, the current-voltage characteristic of each subcell depends only on the gap of the semiconductor, the absorptivity of the cell and whether or not a reflector has been placed at the back of the solar cell. In case a reflector is not provided, it is necessary to know the refraction index of the medium at the back of the cell. Besides this, the cells have been considered to operate at 300 K, illuminated by AM0 spectrum [5] and surrounded by a medium of refraction index equal to unity at the front (air, space).

The modifications of the model we have just referred to are the following:

- The absorptivity of the top subcell has been considered equal to 0.86. This is to allow more light to reach the middle cell and achieve a better current matching between the subcells and, therefore, a higher efficiency.

- For simplicity, no electroluminescent coupling has been considered between the cells. This implies that we have assumed that the solar cells do not emit photons towards the cell located behind and that the cell at the front does not absorb the photons emitted by the cells at its back.

- The degradation of the cells due to electron fluence has been considered by means of the short-circuit current and open-circuit voltage remaining factors. To this end, the values reported in [6] have been assumed. For simplicity, only the GaAs cell has been considered to degrade since, according to these data, this is the cell in the stack that degrades the most. The open-circuit voltage remaining factor has been implemented in practice by multiplying the dark current obtained from the S&Q model by a factor, $f_V$, until the value of the corresponding degraded open-circuit voltage is obtained.

- The solar cell efficiencies thus obtained have been multiplied by a factor $f_\eta$=0.72 to account for other losses such as shadowing factors, non-ideal reflectivity, series resistance, etc. As a result, for example, a beginning of life efficiency for the 3J-2T solar cell of 30 % is obtained, which is a value in the range of the ones reported commercially for this type of cells (Azur Space, for example offers cells with efficiencies between 28 and 32 %) [7].

One consequence of using the model above will be that the differences in efficiencies that will be obtained among the 3J-2T, 3J-6T and 3J-3T can be attributed only to differences in the number of terminals between the cells and not to different assumptions related to radiation degradation which is one of the motivations of this work. We are considering that the implementation of a higher number of terminals, by itself, does not degrade the efficiency of the solar cells. This would be a degradation very difficult to determine and, on the other hand, once we report the results in the next section, the interested readers will be able to easily correct them by the degradation factor they wish. For reference, we plot in Fig. 3 the current-voltage characteristics of the subcells calculated with the model just described. Once the current-voltage characteristics of the subcells are available, obtaining the efficiency for each terminal configuration becomes a trivial electronic circuit exercise and will not be discussed here.

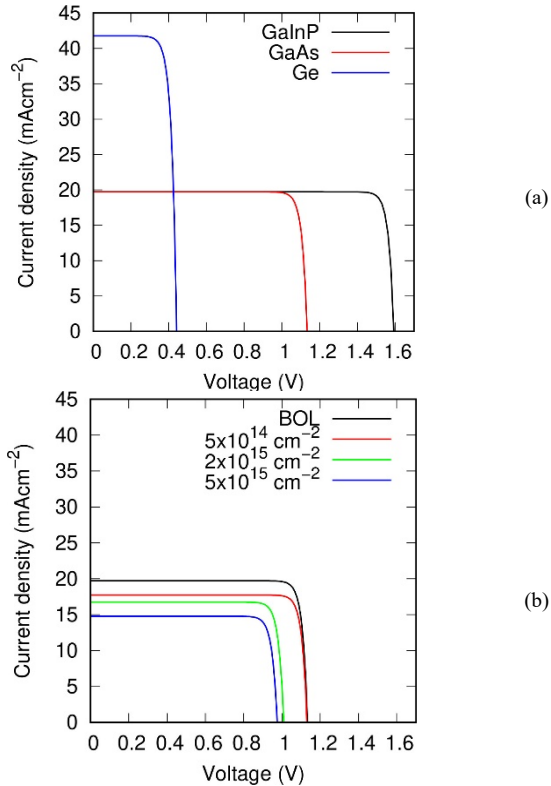

Fig. 2. Current-voltage characteristics of the solar cells used in this paper: (a) characteristics in the radiative limit; (b) degradation assumed of the radiative current-voltage characteristic of the GaAs cell as a function of the 1 MeV electron fluence indicated in the inset. The characteristic in the radiative limit in this plot is labelled BOL. Note that the efficiencies reported in this work correspond to the efficiencies calculated using these curves multiplied by 0.72 (in all cases) to scale the results from the "radiative limit" to results comparable with present state of the art solar cells.

## IV. RESULTS

Fig. 3 summarizes visually the results from the modelling explained before and Table II gives the numerical values. As expected, the most complex structure 3J-6T offers the highest EOL: 29.5 % after a $5 \times 10^{15}$ cm$^{-2}$ fluence, 7.6 points above the EOL of the conventional 3J-2T solar cell which is 21.9 %. However, the 3J-3T structure still offers an EOL of 26.2 % for the same fluence which is 4.3 points above the EOL of the conventional 3J-2T approach while adding only one extra terminal.

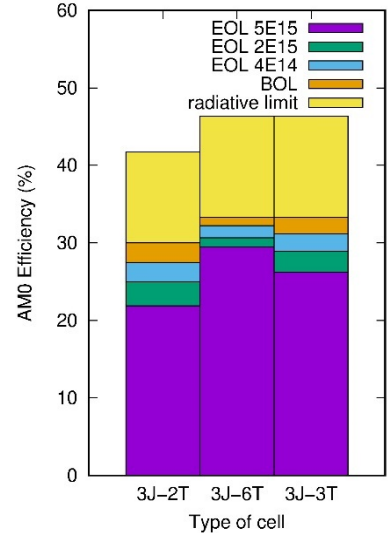

Fig. 3. Efficiency of the three different multi-terminal approaches studied here: 3J-2T (three junctions, two terminals), 3J-6T (three junctions, six terminals), 3J-3T (three junctions, three terminals). Each bar shows the radiative limiting efficiency, the practical efficiency or expected efficiency at the beginning of life (BOL) and the expected end of life efficiency (EOL) after the cells are exposed to the indicated 1 MeV electron fluences (in cm$^{-2}$).

TABLE I. PARAMETERS USED IN THE SIMULATION OF THE SOLAR CELL EFFICIENCY. A REMAINING FACTOR $F_I$ MEANS THAT THE SHORT-CIRCUIT CURRENT IS MULTIPLIED BY $F_I$. A REMAINING FACTOR $f_V$ MEANS THAT THE REVERSE SATURATION CURRENT AT THE BEGINNING OF LIFE (BOL) IS MULTIPLIED BY $f_V$.

|  | InGaP cell | GaAs cell | Ge cell |
|---|---|---|---|
| Spectrum | AM0 (2000 ASTM Standard Extraterrestrial Spectrum Reference E-490-00) | | |
| Absorptivity | 0.86 | 1 | 1 |
| Temperature | 300 K | | |
| Energy gaps | 1.88 eV | 1.41 eV | 0.66 eV |
| Remaining factors for $5 \times 10^{14}$ cm$^{-2}$ 1 MeV electron fluence | Current: $F_I = 1$ Voltage: $f_V = 1$ | Current: $F_I = 0.90$ Voltage: $f_V = 1$ | Current: $F_I = 1$ Voltage: $f = 1$ |
| Remaining factors for $2 \times 10^{15}$ cm$^{-2}$ 1 MeV electron fluence | Current: $F_I = 1$ Voltage: $f_V = 1$ | Current: $F_I = 0.85$ Voltage: $f_V = 106$ | Current: $F_I = 1$ Voltage: $f_V = 1$ |
| Remaining factors for $5 \times 10^{15}$ cm$^{-2}$ 1 MeV electron fluence | Current: $F_I = 1$ Voltage: $f_V = 1$ | Current: $F_I = 0.75$ Voltage: $f_V = 350$ | Current: $F_I = 1$ Voltage: $f_V = 1$ |

TABLE II. EFFICIENCIES OF THE TRIPLE JUNCTION SOLAR CELL DEPENDING ON THE MULTI-TERMINAL CONFIGURATION AND THE ELECTRON FLUENCE.

|  | 3J-2T | 3J-6T | 3J-3T |
|---|---|---|---|
| Radiative limit | 41.8 | 46.4 | 46.4 |
| BOL | 30.0 | 33.3 | 33.3 |
| After $5 \times 10^{14}$ cm$^{-2}$ 1 MeV electron fluence | 27.5 | 32.2 | 31.2 |
| After $5 \times 10^{14}$ cm$^{-2}$ 1 MeV electron fluence | 25.0 | 30.7 | 28.9 |
| After $5 \times 10^{14}$ cm$^{-2}$ 1 MeV electron fluence | 21.9 | 29.5 | 26.2 |

## V. SOME CONSIDERATIONS RELATED TO MULTI-TERMINAL SOLAR CELL INTERCONNECTION IN A MODULE

After the solar cell efficiency analysis we have carried out, we want to remind the reader about some of the topics related to solar cell interconnection in order to implement a photovoltaic module. We will support our discussion with the aid of Fig. 4 that makes a comparison between the connectivity options in a module arrangement between 2-terminal MJSC (Fig. 4a), 4-terminal MJSC (Fig. 4b) and 3-terminal heterojunction bipolar transistor solar cells (Fig. 4c). In all cases, for simplicity, we will consider that the module consists of only two multi-junction solar cells, C1 and C2, being the scalation to a higher number of solar cells trivial. Also, in order to keep the schematics simple, each multi-junction solar cell, C1 and C2, will be assumed to consists of two cells, a top subcell (designated by T) and a bottom subcell (designated by B).

In the case of the two-terminal (2T) multi-junction solar cells (Fig. 4a), the solar cells are usually connected in series in order to produce a high output voltage, $V_H$. As it is known, the drawback of this connection is that the subcells T and B must be current-matched.

In the case of the four-terminal (4T) multi-junction solar cells (Fig. 4b), the top cells can be connected in series to produce a high output voltage $V_{H,T}$. Similarly, the bottom cells, can be connected in series to produce a high output voltage $V_{H,B}$. The drawback of this approach steams from the higher number of terminals that have to be extracted from the multi-junction solar cells and from the fact that the top cell and the bottom cell have to be electrically isolated from each other.

In the case of the three-terminal multi-junction solar cells, there is no trivial way of connecting them in series to increase the output voltage. There is a partial solution, proposed by Gee [8] in 1988 but this solution demands that the output voltage of the top cell is an integer multiple of the output voltage of the bottom cell. Some of the present authors have already explored experimentally this solution in [9] using AlGaAs/GaAs three-terminal heterojunction bipolar transistor solar cells as proof of concept [10].

An alternative to Gee's solution, not demanding any constrain between the output voltages nor the currents of the subcells, could be to connect first the cells in parallel (if needed to increase the current), as illustrated in Fig. 4c, (top cells with top cells and bottom cells with bottom cells) and then, connect each group to a DC/DC converter that directly tracks the maximum power point and regulates and increases the output voltage. The drawback of this approach is that DC/DC converters, to our knowledge, are not very efficient (below 90 %) when their input voltage is too low. As it can be seen, under this approach, while one of the DC/DC converters is fed with the output voltage of the top cell, $V_T$, that is typically above 1 V and could be acceptable, the other DC/DC converter is fed with the output voltage of the bottom cell, $V_B$, which could be too low (below 0.5 V) in order to achieve high efficiency in the DC/DC converter [11]. Hence, the efficient use of three-terminal solar cells is probably linked to the success in the creation of efficient DC/DC boost converters capable of operating with low input voltages.

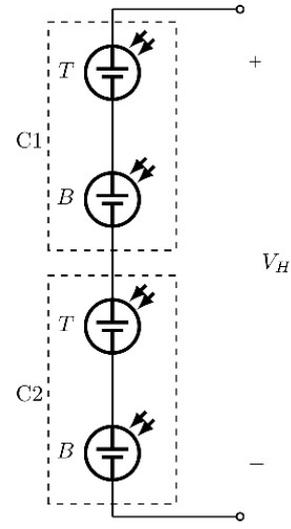

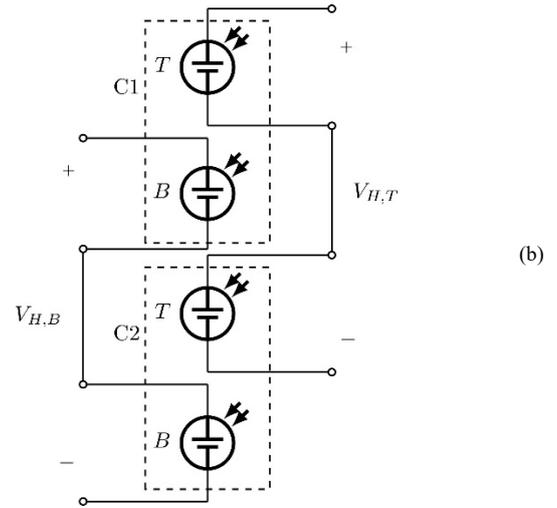

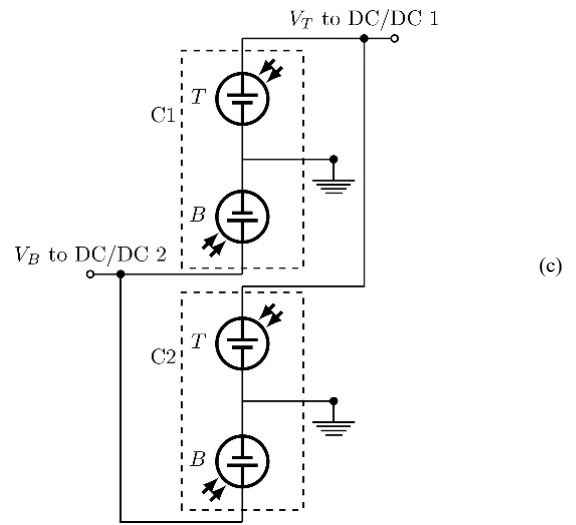

Fig. 4. Examples of interconnection of two multi-junction solar cells (C1 and C2), consisting of two subcells each (top cell, T, and bottom cell, B) aiming to increaset the output voltage: (a) classic series connection of two-terminal solar cells; (b) series connection (grouping top with top cellls and bottom with bottom cells in series) of four-terminal cells and; (c) two three-terminal heterojuntion bipolar transistor solar cells connected in parallel first and then to two DC/DC converters.

(a)

## VI. Conclusions

Considering the multi-junction solar cells before being integrated in a module, the three-terminal approach offers around 4 additional EOL efficiency points when compared with the conventional two-terminal triple-junction solar cell at the cost of introducing one extra terminal but with the advantage of a simpler internal semiconductor layer structure (avoids two tunnel junctions). The six-terminal approach offers 7.6 more efficiency points but at the cost of adding four more terminals and introducing the need of substituting the tunnel junctions by electrically isolating layers and, therefore, not simplifying the internal semiconductor structure of the multi-junction solar cell. We point out though, that due to difficulty for interconnecting the three-terminal solar cells in series to increase the voltage, their success is linked to the success of the development of efficient DC/DC boost converters capable of operating with low input voltages (< 0.5 V) in order to boost the output voltage of the bottom cell to useful levels.

## References

[1] J. Villa and A. Marti, "Impact of the Spectrum in the Annual Energy Production of Multijunction Solar Cells," *IEEE J. Photovoltaics*, vol. 7, no. 5, pp. 1479–1484, Sep. 2017.

[2] "Shockley, Queisser - 1961 - Detailed Balance Limit of Efficiency of p-n Junction Solar Cells.pdf," J. Appl. Phys., vol. 32, no. 3, pp. 510–519, 1961.

[3] G. L. Araújo and A. Martí, "Absolute limiting efficiencies for photovoltaic energy conversion," Sol. Energy Mater. Sol. Cells, vol. 33, no. 2, pp. 213–240, Jun. 1994.

[4] A. Martí and G. L. Araújo, "Limiting efficiencies for photovoltaic energy conversion in multigap systems," Sol. Energy Mater. Sol. Cells, vol. 43, no. 2, pp. 203–222, Sep. 1996.

[5] N. Webpage, "2000 ASTM Standard Extraterrestrial Spectrum Reference E-490-00." [Online]. Available: https://rredc.nrel.gov/solar//spectra/am0/. [Accessed: 13-Apr-2009].

[6] R. Campesato et al., "31% EUROPEAN INGAP/GAAS/INGANAS SOLAR CELLS FOR SPACE APPLICATION," 6th IASS Conf., vol. 16, 2017.

[7] "SPACE Solar Cells (AZUR SPACE Solar Power GmbH)." [Online]. Available: http://www.azurspace.com/index.php/en/products/products-space/space-solar-cells. [Accessed: 29-Aug-2019].

[8] J. M. Gee, "A comparison of different module configurations for multi-band-gap solar cells," Sol. Cells, vol. 24, no. 1–2, pp. 147–155, May 1988.

[9] M. Zehender et al., "Module interconnection for the three-terminal heterojunction bipolar transistor solar cell," AIP Conf. Proc., vol. 2012, no. 1, p. 64189, Sep. 2018.

[10] A. Martí and A. Luque, "Three-terminal heterojunction bipolar transistor solar cell for high-efficiency photovoltaic conversion," Nat. Commun., vol. 6, pp. 6902–6902, Apr. 2015.

[11] M. Pollak, L. Mateu, and P. Spies, "Step-Up DC-DC-Converter With Coupled Inductors for Low Input Voltages," Proc. PowerMEMS 2008, no. January 2008, pp. 145–148, 2008.

This work was supported by ESA though the Contract No. 4000127130/19/NL/FE. There has been also partial support from the Project GRECO (787289 H2020 Grant Agreement), funded by the European Commission, that ensures this work is aligned to the European Open Science Policy. E. A. is funded by a Ramón y Cajal Fellowship from the Spanish Ministry of Science (RYC-2015-18539). M.H.Z. is grateful to the Universidad Politécnica de Madrid for the funding from the 'Programa Propio para Ayudas Predoctorales de la UPM'.